# Load-Modulated Single-RF MIMO Transmission for Spatially Multiplexed QAM Signals

Seung-Eun Hong
Wireless Transmission Research Dep.
Commun. & Internet Research Lab., ETRI
Daejeon, Korea
iptvguru@etri.re.kr

Kyoung-Sub Oh

Gamma Nu, Inc.
Hwaseong, Korea
ksoh@gammanu.com

*Abstract*—Today, MIMO has become an indispensable scheme for providing significant spectral efficiency in wireless communication and for future wireless system, recently, it goes to two extremes: massive MIMO and single-RF MIMO. This paper, which is put in the latter, utilizes load-modulated arrays with only reactance loads for single-RF transmission of spatially multiplexed QAM signals. To alleviate the need for iterative processes while considering mutual coupling in the compact antenna, we present a novel design methodology for the loading network, which enables the exact computation of the three reactance loads per antenna element and also the perfect matching to the source with the opportunity to select appropriate analog tunable loads. We verify the design methodology by comparing the calculated values for some key parameters with the values from the circuit simulation. In addition, as an evaluation of the proposed architecture, we perform the bit error rate (BER) comparison which shows that our scheme with ideal loading is comparable to the conventional MIMO.

## I. INTRODUCTION

Even though MIMO systems have been well known to be able to provide significant spectral efficiency, the deployment is really limited due to the multiple radio-frequency (RF) chains, widely spaced antenna elements and power consumption [1]. To find a way out of the aforementioned problems, electronically steerable parasitic antenna radiators (ESPARs) with compactness have been considered to provide spatial multiplexing MIMO only with a single-RF chain and purely imaginary loads over beam-space domain. Recently, ESPARs with complex loading have been proposed for single-RF MIMO transmission of higher order modulations [2]-[4]. The recent works need an active circuit which implements a tunable complex loading for the parasitic elements with the resistance ranging from negative to positive values. However, the ESPAR-MIMO with the complex loading results in severe power consumption due to the active load for generating negative real impedance and needs high-capability circulator in order to prevent the power amplifier in the single-RF chain from being broken down since the power returns back into the source side. In addition, the active circuit for tunable complex loading might be oscillated since the power also puts into the active circuit. It is also noted that the preceding works do not address the phase error in transmitting spatially multiplexed symbols [5].

More recently, in order to solve some issues in the cost and the size of massive MIMO transmitters, a new type of MIMO transmitter with a single-RF chain called LMA (Load Modulated Array) has been proposed [6]-[8]. All of works, however, assume no mutual coupling between the LMA elements and therefore the design of the load modulators is enabled with the simple and analytical equations in [6]. The assumption is not practical in real environment such as small base station and terminal with the emphasis of compact size.

Thus, the main contributions of this paper are three-fold: 1) for single-RF LMA-MIMO transmission of spatially multiplexed QAM signals, we present a methodology for loading network which enables the exact computation of the reactance loads and also the self-satisfied impedance matching with opportunity to select appropriate analog tunable loads for various passive loading circuits; 2) we address the solution to the phase error and the amplitude adjustment in transmitting spatially multiplexed symbols by providing closed-form equations; and 3) as an evaluation of the proposed scheme, we verify the design methodologies by comparing the calculated values for some key parameters with the values from the circuit simulation and also present BER comparisons where the proposed LMA-MIMOs with ideal loading and some erred loadings are compared with conventional MIMO transmitter.

The rest of the paper will be organized as follows: Section II describes an appropriate design concept that can solve the addressed problems. In Section III, we identify some design parameters for the proposed system and derive the closed-form equations to determine the values of the design parameters. Section IV presents the numerical and simulation results. The concluding remarks are addressed in Section V.

## II. SPATIAL MULTIPLEXING OF 16-QAM SIGNALS WITH LMA

### A. Beamspace MIMO Multiplexing

In the context of the single-RF MIMO transmitter, each of spatially multiplexed streams is mapped into the respective basis beam pattern, which is contrast to conventional MIMO transmitters where symbol streams are driven to different antenna elements. The radiated beam pattern of this beam-

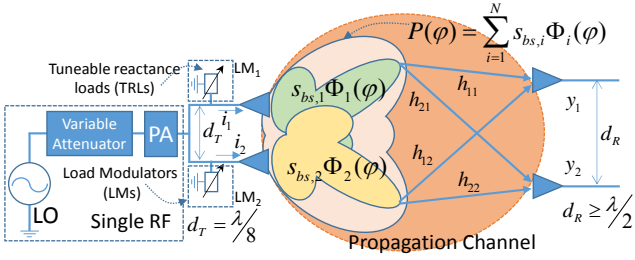

Fig. 1. Load-modulated (LM) beam-space MIMO architecture

space multiplexing with a two-element antenna can be decomposed as [2],[9]:

$$P_T(\theta,\varphi) = s_{bs,1}\Phi_1(\theta,\varphi) + s_{bs,2}\Phi_2(\theta,\varphi)$$
$$= (i_1 q_{11} + i_2 q_{21})\Phi_1(\theta,\varphi) + (i_1 q_{12} + i_2 q_{22})\Phi_2(\theta,\varphi) \quad (1)$$

where $q_{mn}$ denotes the projection of $m^{th}$ element in the antenna steering vector onto the $n^{th}$ orthonormal function ($\Phi_n$), $i_m$ denotes the current to $m^{th}$ antenna port, and $s_{bs,n}$ is one of spatially multiplexed symbols which weights or modulates the $n^{th}$ basis pattern. Based on [3], the currents generated for the beamspace multiplexing signals in (1) are given as:

$$\begin{bmatrix} i_1 \\ i_2 \end{bmatrix} = \begin{bmatrix} 1/\sqrt{2\pi} & -I_0(jb)/\left(\sqrt{2\pi}\sqrt{1-I_0^2(jb)}\right) \\ 0 & 1/\left(\sqrt{2\pi}\sqrt{1-I_0^2(jb)}\right) \end{bmatrix} \begin{bmatrix} \hat{s}_{bs,1} \\ \hat{s}_{bs,2} \end{bmatrix} \quad (2)$$

where $I_0(jb)$ is the zero$^{th}$ order modified Bessel function of the first kind and instead of using the beamspace multiplexing signals we use power-normalized and phase-shifted signals which are given by

$$\hat{s}_{bs,i} = q e^{j\phi} s_{bs,i}, \quad i=1,2. \quad (3)$$

With the Z-parameter for the two-element antenna, the radiated power is calculated as follows:

$$P_{rad} = \text{Re}\left(v_1 i_1^* + v_2 i_2^*\right) = \text{Re}\left(Z_{11} i_1 i_1^* + Z_{12} i_2 i_1^* + Z_{21} i_1 i_2^* + Z_{22} i_2 i_2^*\right)$$
$$= \frac{|s_{bs,1}|^2 + |s_{bs,2}|^2}{|s_{bs,1}|_{max}^2 + |s_{bs,2}|_{max}^2} P_{rad,max} \quad (4)$$

where the maximum radiated power $P_{rad,max}$ is assumed to have 1W without loss of generality. The power normalization factor q is determined to make the radiated power of the beamspace multiplexing signals $P_{rad}$ in (4).

### B. Load-Modulation based Single-RF MIMO Transmission

The concept of load-modulation based single-RF MIMO transmission is illustrated in Fig. 1, where the LMA elements are separated by λ/8 and branched from the single-RF chain. The transmitting signals are mapped onto the basis patterns and the baseband signals can be retrieved successfully by using a conventional MIMO receiver, like [2].

With the load modulation, the single-RF chain consists of only a simple oscillator which provide a sinusoid of fixed amplitude and phase, and an attenuator and a power amplifier which control the radiated power according to (4). In other words, the RF circuit requires no mixer and no voltage modulation unlike ESPAR. In addition, our load modulation scheme does not suffer from the impedance mismatch problem of the circuit to the antenna impedance since load modulators (LMs) are sophisticatedly designed to have the perfect matching to the source. The load modulators (LM$_1$ and LM$_2$) are lossless two-port networks. The LMs can be implemented by a T-type or Π-type with/without transmission line. In the sequel, the design procedure of the proposed transmitter is described.

### III. LM Design and Computational Framework

In this section, with two-element LMA antenna, we make an effort to build an analytical framework that enables us to obtain the sets of reactance loading values which exploit to the maximum multiplexing capabilities of the given LMA. We consider that two streams of 16-QAM signals are transmitted with the LMA.

#### A. LM with Tuneable Reactance Loads

Various passive loading circuits for our load modulators are shown in Fig. 2. As shown in the figure, three TRLs are required for LM per antenna element and a transmission line (TL) may be optionally used. The TRLs can be formed with T-type or Π-type. As a specific example, in this paper, we consider a T-type LM with the TL.

Through a well-known ABCD parameter representation, the two-port networks for LM$_i$ shown in Fig. 2 are given by

$$\begin{bmatrix} V_i' \\ I_i' \end{bmatrix} = \begin{bmatrix} a_i & jZ_0 b_i \\ j\frac{1}{Z_0} c_i & d_i \end{bmatrix} \begin{bmatrix} V_i \\ I_i \end{bmatrix} \quad (5)$$

where $I_i$ is equivalent to $i_i$ in (2) and $a_i$, $b_i$, $c_i$, $d_i$ are represented with the normalized reactance tuples ($x_{1i}$, $y_{2i}$, $x_{3i}$), $i = \{1,2\}$, in T-type with TL as follow:

$$a_i = -y_{2i}, \quad b_i = 1 - x_{3i} y_{2i}, \quad (6)$$
$$c_i = 1 - x_{1i} y_{2i}, \quad d_i = -x_{3i} - x_{1i}(1 - x_{3i} y_{2i})$$
$$x_{1i} = X_{1i}/Z_0, \quad y_{2i} = Z_0 Y_{2i}, \quad x_{3i} = X_{3i}/Z_0. \quad (7)$$

In case of Π-type without TL, the parameters mapping in (6) is given by

$$a_i = -y_{3i} - y_{1i}(1 - x_{2i} y_{3i}), \quad b_i = 1 - x_{2i} y_{1i}, \quad (6')$$
$$c_i = 1 - x_{2i} y_{3i}, \quad d_i = -x_{2i}$$

The tuples ($X_{1i}$, $Y_{2i}$, $X_{3i}$), $i = \{1,2\}$, are values of TRLs and $Z_0$ is source impedance of 50 Ω. From (6), we can see that those are related with

$$d_i = \frac{1 - b_i c_i}{a_i}. \quad (8)$$

Like the normalized reactance tuples, we use normalized Z-parameters for the antenna as follow:

$$(Z_{mn})_{m,n\in\{1,2\}} = (Z_0(z_{mn,r} + jz_{mn,i})) \quad (9)$$

where $Z_{mn}$ is the mutual impedance between the $m$th and the $n$th port in the antenna array with 2 elements viewed as an 2-port microwave network [11], and $z_{mn,r}$ ($z_{mn,i}$) represents normalized real(imaginary) component of $Z_{mn}$.

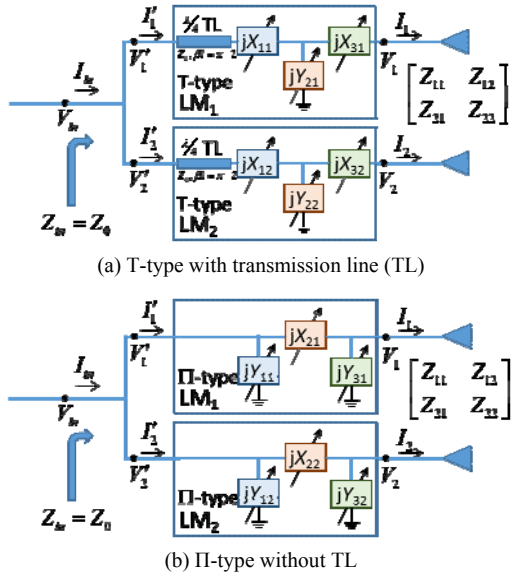

(a) T-type with transmission line (TL)

(b) Π-type without TL

Fig. 2. Pictorial examples of various pasive loading circuits for LMs

### B. Design Procedure 1: Computation of TRLs

For the design goal of avoiding that a dynamic matching network is needed, we design LMs so that the input impedance $Z_{in}$ from Fig. 2 can be automatically matched to $Z_0$. For that purpose, firstly, by applying Kirchoff's circuit law at the input port in the LM single-RF MIMO system and simplifying in terms of $I_1$ by introducing $\eta = \eta_r + j\eta_i = I_2/I_1$, we can derive these equations:

$$V_{in} = V_1' = Z_0\left[\delta_{1-}a_1 + j(\delta_{1+}a_1 + b_1)\right]I_1$$
$$V_{in} = V_2' = Z_0\left[(\delta_{2-}a_2 - \eta_i b_2) + j(\delta_{2+}a_2 + \eta_r b_2)\right]I_1 \quad (10)$$
$$I_{in} = \left[(d_1 + \eta_r d_2 - \delta_{1+}c_1 - \delta_{2+}c_2) + j(\eta_i d_2 + \delta_{1-}c_1 + \delta_{2-}c_2)\right]I_1$$

where

$$\delta_{1-} = z_{11,r} + \eta_r z_{12,i} - \eta_i z_{12,r},\ \delta_{1+} = z_{11,i} + \eta_r z_{12,r} + \eta_i z_{12,i}$$
$$\delta_{2-} = z_{21,r} + \eta_r z_{22,r} - \eta_i z_{22,i},\ \delta_{2+} = z_{21,i} + \eta_r z_{22,i} + \eta_i z_{22,r} \quad (11)$$

Due to $V_{in} = V_1' = V_2'$, the following relationship can be derived from (10):

$$\delta_{1-}a_1 - \delta_{2-}a_2 = -\eta_i b_2$$
$$\delta_{1+}a_1 - \delta_{2+}a_2 = \eta_r b_2 - b_1 \quad (12)$$

Defining additionally $\Delta_{12} = \delta_{1+}\delta_{2-} - \delta_{1-}\delta_{2+}$, $\Omega_{21} = \eta_r\delta_{1-} + \eta_i\delta_{1+}$ and $\Omega_{22} = \eta_r\delta_{2-} + \eta_i\delta_{2+}$, we can derive two relations from (12) as follow:

$$a_1 = \frac{\Omega_{22}b_2 - \delta_{2-}b_1}{\Delta_{12}},\ a_2 = \frac{\Omega_{21}b_2 - \delta_{1-}b_1}{\Delta_{12}} \quad (13)$$

Now, for the design goal of the self-satisfied impedance matching that the input impedance $Z_{in}$ should be equal to the $Z_0$, the condition is derived as follows:

$$\delta_{1-}a_1 = d_1 + \eta_r d_2 - \delta_{1+}c_1 - \delta_{2+}c_2$$
$$\delta_{1+}a_1 + b_1 = \eta_i d_2 + \delta_{1-}c_1 + \delta_{2-}c_2 \quad (14)$$

Replacing $d_1$ and $d_2$ in (14) with (8) results in a system of two equations involving $c_1$ and $c_2$, which is given by

$$\left(\delta_{1+} + \frac{b_1}{a_1}\right)c_1 + \left(\delta_{2+} + \eta_r\frac{b_2}{a_2}\right)c_2 = \frac{1}{a_1} + \eta_r\frac{1}{a_2} - \delta_{1-}a_1$$
$$\delta_{1-}c_1 + \left(\delta_{2-} - \eta_i\frac{b_2}{a_2}\right)c_2 = \delta_{1+}a_1 + b_1 - \eta_i\frac{1}{a_2} \quad (15)$$

It is noted that the two equations in (15) are not independent and thus, to find a solution, the following relation should be satisfied:

$$\delta_{1-}\left(\frac{1}{a_1} + \eta_r\frac{1}{a_2} - \delta_{1-}a_1\right) = \left(\delta_{1+} + \frac{b_1}{a_1}\right)\left(\delta_{1+}a_1 + b_1 - \eta_i\frac{1}{a_2}\right) \quad (16)$$

By replacing $a_1$ and $a_2$ in (16) with (13), we can derive a quadratic equation with unknowing variable $b_2$ so that it is solved in terms of $b_1$ as follows:

$$b_2 = b_1 \frac{\delta_{1-}(\delta_{1-}\delta_{2-} + \delta_{1+}\delta_{2+})}{(\delta_{1+}^2 + \delta_{1-}^2)\Omega_{22}} \times$$
$$\left\langle 1 \pm \left|\frac{\Delta_{12}}{\delta_{1+}\delta_{2+} + \delta_{1-}\delta_{2-}}\right|\sqrt{\frac{(\delta_{1+}^2 + \delta_{1-}^2)}{\delta_{1-}^2 b_1^2}(\delta_{1-} + \Omega_{22}) - 1}\right\rangle \quad (17)$$

Therefore, if we select a value of $b_1$ as a free variable then we can consecutively find $b_2$, $a_1$ and $a_2$. In addition, from the second equation in (15), if we select a value of $c_2$ as another free variable then we can also find $c_1$, vice versa. In other words, we can take two design parameters of $b_1$ and $c_2$ (or $c_1$) for selecting TRLs.

### C. Design Procedure 2: Refinement of the Computation

We start this sub-section by considering phase component of the given current. From (10), we can see the following phase relationship

$$\angle \frac{I_1}{I_{in}} = \angle \frac{Z_0 I_1}{V_{in}} = \angle \frac{I_1}{V_{in}} = \tan^{-1}\left(\frac{-\delta_{1+}a_1 - b_1}{\delta_{1-}a_1}\right) = \theta \quad (18)$$

In addition, (2) says that the phase of $I_2$ is determined only by $s_{bs,2}$. Therefore, from the understandings of (18) and (2), we can derive the relationship of phases as follows:

$$\theta_{\angle V_{in}=0} = \angle I_1 = \angle I_2 + \angle\left(\frac{I_1}{I_2}\right) + \angle s_{bs,2} - \angle s_{bs,2}$$
$$= \angle I_2 - \angle s_{bs,2} + \angle s_{bs,2} - \angle \eta = \phi + \angle s_{bs,2} - \angle \eta \quad (19)$$

By replacing $a_1$ from (18) with (13), we can get

$$\Omega_{22}(\delta_{1+} + \delta_{1-}\tan\theta)b_2 = \delta_{1-}(\delta_{2+} + \delta_{2-}\tan\theta)b_1 \quad (20)$$

By substituting and simplifying with (17) and (20), $b_1$ and $b_2$ can be written as follows:

$$b_1 = \pm\frac{\delta_{1+} + \delta_{1-}\tan\theta}{\delta_{1-}\sqrt{1+\tan^2\theta}}\sqrt{\delta_{1-} + \Omega_{22}},\ b_2 = \pm\frac{\delta_{2+} + \delta_{2-}\tan\theta}{\Omega_{22}\sqrt{1+\tan^2\theta}}\sqrt{\delta_{1-} + \Omega_{22}} \quad (21)$$

Thus, the remaining parameters are given by

$$a_2 = \mp\frac{\eta_r - \eta_i\tan\theta}{\Omega_{22}\sqrt{1+\tan^2\theta}}\sqrt{\delta_{1-} + \Omega_{22}},\ a_1 = \mp\frac{1}{\delta_{1-}\sqrt{1+\tan^2\theta}}\sqrt{\delta_{1-} + \Omega_{22}} \quad (22)$$

Then, if we substitute the corresponding parameters of the second equation in (15) with (21)~(22) and introduce a new parameter $s$, we can drive two equations as follow:

$$\frac{\Omega_{22}}{\eta_r - \eta_i \tan\theta}\left(c_2 \mp \frac{\eta_i + \eta_r \tan\theta}{\sqrt{1+\tan^2\theta}\sqrt{\delta_{1-}+\Omega_{22}}}\right) = s, \quad (23)$$

$$\delta_{1-}\left(c_1 \mp \frac{\tan\theta}{\sqrt{1+\tan^2\theta}\sqrt{\delta_{1-}+\Omega_{22}}}\right) = -s. \quad (24)$$

Thus, we can compute $c_1$ and $c_2$ with arbitrary $s$ as follow:

$$c_1 = -\frac{1}{\delta_{1-}}s \pm \frac{\tan\theta}{\sqrt{1+\tan^2\theta}\sqrt{\delta_{1-}+\Omega_{22}}}, \quad (25)$$

$$c_2 = \frac{\eta_r - \eta_i \tan\theta}{\Omega_{22}}s \pm \frac{\eta_i + \eta_r \tan\theta}{\sqrt{1+\tan^2\theta}\sqrt{\delta_{1-}+\Omega_{22}}}. \quad (26)$$

Although we are able to compute the TRLs with (21)~(22) and (25)~(26), all of the given equations has plus-minus sign so that we have to decide the sign. For the purpose, we revisit (21) and (22) with $\tan\theta = \sin\theta/\cos\theta$.

$$b_1 = \pm\frac{\delta_{1+}\cos\theta + \delta_{1-}\sin\theta}{\delta_{1-}}\frac{\sqrt{\delta_{1-}+\Omega_{22}}}{\mathrm{sgn}(\cos\theta)} \quad (21')$$

$$a_1 = \mp\frac{\cos\theta}{\delta_{1-}}\frac{\sqrt{\delta_{1-}+\Omega_{22}}}{\mathrm{sgn}(\cos\theta)} \quad (22')$$

Also revisiting (18) and using trigonometric ratio in the right triangle, we can get $\cos\theta$ and $\sin\theta$ as follow:

$$\cos\theta = \frac{\delta_{1-}a_1}{\sqrt{(\delta_{1-}a_1)^2+(\delta_{1+}a_1+b_1)^2}}, \quad \sin\theta = \frac{-(\delta_{1+}a_1+b_1)}{\sqrt{(\delta_{1-}a_1)^2+(\delta_{1+}a_1+b_1)^2}}. \quad (18')$$

By substituting the corresponding parameters in (18′) with (21′) and then simplifying it, we can get the relation given by

$$\cos\theta = \mp\frac{\cos\theta}{\mathrm{sgn}(\cos\theta)}, \quad \sin\theta = \mp\frac{\sin\theta}{\mathrm{sgn}(\cos\theta)} \quad (27)$$

From (27), if the sign of $\cos\theta$ is minus, i.e., $\mathrm{sgn}(\cos\theta) = -1$, then we have to select the minus sign in (27). In other words, the sign in (27) can be selected with same as the sign of $\cos\theta$. Thus, those parameters of (21)~(22), (25)~(26) are simplified like these:

$$b_1 = -\frac{\delta_{1+}\cos\theta+\delta_{1-}\sin\theta}{\delta_{1-}}\sqrt{\delta_{1-}+\Omega_{22}}, b_2 = -\frac{\delta_{2+}\cos\theta+\delta_{2-}\sin\theta}{\Omega_{22}}\sqrt{\delta_{1-}+\Omega_{22}}$$

$$a_1 = \frac{\cos\theta}{\delta_{1-}}\sqrt{\delta_{1-}+\Omega_{22}}, \quad a_2 = \frac{\eta_r\cos\theta - \eta_i\sin\theta}{\Omega_{22}}\sqrt{\delta_{1-}+\Omega_{22}} \quad (28)$$

$$c_2 = \frac{\eta_r - \eta_i\tan\theta}{\Omega_{22}}s - \frac{\eta_i\cos\theta+\eta_r\sin\theta}{\sqrt{\delta_{1-}+\Omega_{22}}}, \quad c_1 = -\frac{1}{\delta_{1-}}s - \frac{\sin\theta}{\sqrt{\delta_{1-}+\Omega_{22}}}.$$

### D. Power Normalization of Beamspace Multiplexed Signals

The power of radiated beam carrying the beamspace signals pair should be normalized to be correctly distributed so that the beamspace multiplexed signals are appropriately scattered in their constellation diagram. To the end, we use power-normalized and phase-shifted signals, as shown in (3), and the resultant currents are given by

$$I_1 = \frac{qe^{j\phi}}{\sqrt{2\pi}}\left(s_{bs,1} - \frac{I_0(jb)}{\sqrt{1-I_0^2(jb)}}s_{bs,2}\right), \quad I_2 = \frac{qe^{j\phi}}{\sqrt{2\pi}}\frac{1}{\sqrt{1-I_0^2(jb)}}s_{bs,2} \quad (29)$$

where the power-normalization factor $q$ is used to make the radiated beam power be equal to (4) and the value is given by (30) due to the assumption of $P_{rad,max}=1W$ and the relation as follows:

$$P_{rad} = \frac{|V_{in}|^2}{Z_0} = \frac{|s_{bs,1}|^2+|s_{bs,2}|^2}{|s_{bs,1}|^2_{max}+|s_{bs,2}|^2_{max}}. \quad (31)$$

About how to normalize the power of radiated beam, we use a scheme as shown in Fig. 1. More specifically, a simple oscillator is used to provide a sinusoid of fixed amplitude and phase and both an attenuator and a power amplifier are used to control the radiated beam power. Resultantly, the feeding voltage can be controlled as a function of the beamspace multiplexed vector of signals $[s_{bs,1}\ s_{bs,2}]$:

$$|V_{in}| = \sqrt{\frac{|s_{bs,1}|^2+|s_{bs,2}|^2}{|s_{bs,1}|^2_{max}+|s_{bs,2}|^2_{max}}Z_0}.$$

### IV. PERFORMANCE EVALUATION

As an evaluation of the proposed architecture, Table I shows the first 16 computed voltages, currents, phases and powers for the T-type with TL in case of 2-stream multiplexing of 16-QAM signals with LMAs. The Z-parameter for the 2-element LMA antenna is obtained by the CST software and the values are as $Z_{11}=Z_{22}=46.09-j12.18\ \Omega$; $Z_{12}=Z_{21}=18.36-j29.92\ \Omega$. From (2), (3) and (19), the phase $\phi$ should be constant (for an example, we set the value of phase shift in (3) to -10), which is justified by Table I. The sum of the two powers ($P_1$ and $P_2$) which are effective real input powers into two LMA antenna element in the proposed single-RF MIMO is equal to $P_{in}$ and the value is given by

$$P_{in} = \mathrm{Re}(V_{in}I_{in}^*) = Z_0|I_1|^2\left[(\delta_{1-}a_1)^2+(\delta_{1+}a_1+b_1)^2\right] = Z_0|I_1|^2[\delta_{1-}+\Omega_{22}]. \quad (20)$$

$P_1$ and $P_2$ are calculated from $P_{in}$ as follow:

$$P_1 = \frac{\delta_{1-}}{\delta_{1-}+\Omega_{22}}P_{in}, \quad P_2 = \frac{\Omega_{22}}{\delta_{1-}+\Omega_{22}}P_{in} \quad (21)$$

It is mentioned that some negative powers in $P_1$ and $P_2$ cannot cause any problem in our proposed system because the sum powers (i.e. $P_{in}$) are always positive, meaning that the minus power in one side does not return back into the source side but flow into the other passive loading side. The values of sum powers are fully radiated into LMA antenna through lossless LMs and we can see that the radiated powers are well mapped into the constellations of the multiplexed signals.

Fig. 3 gives some simulation results by the CST microwave software. Although we do not describe the detailed explanation due to the limited room of the paper, we can see that the

$$q = \frac{\sqrt{2\pi}}{\sqrt{z_{11,r}|s_{bs,1}|^2 + \frac{z_{11,r}I_0^2(jb)-2z_{21,r}I_0(jb)+z_{22,r}}{|1-I_0^2(jb)|}|s_{bs,2}|^2 - 2\frac{z_{11,r}I_0(jb)-z_{21,r}}{\sqrt{1-I_0^2(jb)}}\mathrm{Re}(s_{bs,1}^*s_{bs,2})}}\frac{|V_{in}|}{Z_0} \quad (30)$$

TABLE I. COMPUTED VOLTAGE, CURRENT, PHASE AND POWER IN T–TYPE WITH TL LM CIRCUIT FOR OUR PROPOSED SYSTEM

| ind. | $V_{in}$ | $I_1$ | $I_2$ | $\phi$ | $P_1$(W) | $P_2$(W) |
|---|---|---|---|---|---|---|
| 1 | 6.236 | 0.060-0.043j | -0.079+0.113j | -10 | 0.173 | 0.605 |
| 2 | 5.27 | 0.072+0.012j | -0.095+0.051j | -10 | 0.28 | 0.275 |
| 3 | 5.27 | 0.073+0.061j | -0.094-0.014j | -10 | 0.421 | 0.135 |
| 4 | 6.236 | 0.071+0.098j | -0.091-0.064j | -10 | 0.569 | 0.209 |
| 5 | 5.27 | 0.006-0.039j | -0.018+0.118j | -10 | -0.015 | 0.571 |
| 6 | 4.082 | 0.022+0.032j | -0.040+0.057j | -10 | 0.166 | 0.167 |
| 7 | 4.082 | 0.026+0.082j | -0.041-0.029j | -10 | 0.36 | -0.026 |
| 8 | 5.27 | 0.028+0.107j | -0.040-0.074j | -10 | 0.461 | 0.095 |
| 9 | 5.27 | -0.055-0.026j | 0.054+0.100j | -10 | -0.053 | 0.609 |
| 10 | 4.082 | -0.053+0.041j | 0.050+0.035j | -10 | 0.068 | 0.266 |
| 11 | 4.082 | -0.031+0.086j | 0.027-0.038j | -10 | 0.276 | 0.058 |
| 12 | 5.27 | -0.017+0.111j | 0.012-0.081j | -10 | 0.416 | 0.14 |
| 13 | 6.236 | -0.096-0.013j | 0.104+0.073j | -10 | 0.063 | 0.715 |
| 14 | 5.27 | -0.088+0.038j | 0.094+0.014j | -10 | 0.135 | 0.421 |
| 15 | 5.27 | -0.072+0.080j | 0.076-0.041j | -10 | 0.276 | 0.279 |
| 16 | 6.236 | -0.058+0.114j | 0.061-0.086j | -10 | 0.452 | 0.326 |

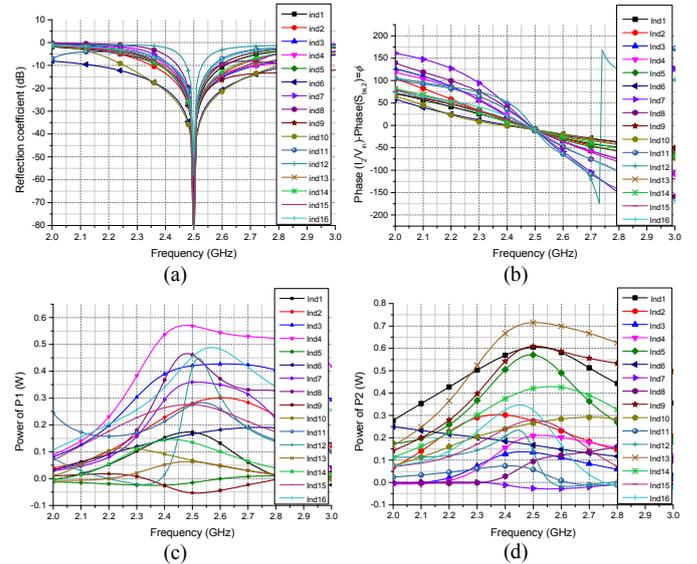

Fig. 3. Simulation results of the proposed T-type with TL LM circuit: (a) Reflection Coefficient; (b) $\phi$; (c) Power of $P_1$; (d) Power of $P_2$

simulation results at the operating frequency 2.5 GHz coincide with the calculated values in Table I. Fig. 4 shows BER performance by the system simulation where the proposed architecture with T-type with TL LMs is compared with conventional MIMO transmitter under the same environment and conditions as in [2]. The results in Fig. 4 indicate that the performance of the LMA-MIMO, using ideal loading values, coincides well with the one obtained by conventional MIMO. We leave the details to our journal version, for the limitation of the space in this paper.

## V. CONCLUDING REMARKS

In this paper, we have presented a design methodology for a single-RF LMA-MIMO transmission of spatially multiplexed QAM signals. To alleviate the need for iterative processes while considering mutual coupling in the compact antenna, the design methodology for the loading network enables us to calculate the exact values of the three reactance loads per antenna element and to automatically achieve the perfect matching into the source. We verify the design methodology by comparing the calculated values for some key parameters with the values from the circuit simulation. In addition, as an evaluation of the proposed architecture, we perform the BER comparison where the proposed LMA-MIMOs with ideal loading and some erred loadings are compared with conventional MIMO transmitter and we show that the performance of the LMA-MIMO, using ideal loading values, coincides well with the one obtained by conventional MIMO.

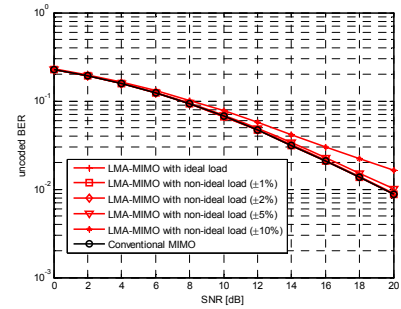

Fig. 4. Comparison of BER performance